# THz-driven ultrafast spin-lattice scattering in amorphous metallic ferromagnets

S. Bonetti[1]*, M.C. Hoffmann[2], M.-J. Sher[3,4], Z. Chen[3,5], S.-H. Yang[6], M. Samant[6], S.S.P. Parkin[6,7], H.A. Dürr[3]*

[1] Department of Physics, Stockholm University, Stockholm, Sweden

[2] Linac Coherent Light Source, SLAC National Accelerator Laboratory, 2575 Sand Hill Road, Menlo Park, CA 94025, USA.

[3] Stanford Institute for Materials and Energy Sciences, SLAC National Accelerator Laboratory, 2575 Sand Hill Road, Menlo Park, CA 94025, USA.

[4] Department of Applied Physics, Stanford University, Stanford, California 94305, USA

[5] Department of Physics, Stanford University, Stanford, California 94305, USA

[6] IBM Almaden Research Center, San Jose CA 95120, USA

[7] Max-Planck Institut für Mikrostrukturphysik, Weinberg 2, Halle, Germany

*email: stefano.bonetti@fysik.su.se; hdurr@slac.stanford.edu

**Abstract**

We use single-cycle THz fields and the femtosecond magneto-optical Kerr effect to respectively excite and probe the magnetization dynamics in two thin-film ferromagnets with different lattice structure: crystalline Fe and amorphous CoFeB. We observe Landau-Lifshitz-torque magnetization dynamics of comparable magnitude in both systems, but only the amorphous sample shows ultrafast demagnetization caused by the spin-lattice depolarization of the THz-induced ultrafast spin current. Quantitative modelling shows that such spin-lattice scattering events occur on similar time scales than the conventional spin conserving electronic scattering (~30 fs). This is significantly faster that optical laser-induced demagnetization. THz conductivity measurements point towards the influence of lattice disorder in amorphous CoFeB as the driving force for enhanced spin-lattice scattering.

The interaction between magnetism and light is receiving considerable interest after the groundbreaking experiments that showed sub-ps demagnetization of ferromagnets by fs optical laser [1] or even ultrafast all-optical magnetic switching in magnetic alloys [2]. Subsequent to these pioneering studies, these phenomena have been observed in a variety of different materials and experimental conditions [3–20], creating much interest in the possibility of realizing ultrafast magnetic data storage controlled by light. However, even to date, the fundamental physical processes governing the ultrafast magnetization remains debated within the scientific community.

A particularly controversial issue is related to the microscopic mechanism of how a ferromagnet is able to dissipate its spin angular momentum at the sub-ps time scale. Angular momentum carried by the optical laser pulses was ruled out as being orders of magnitude too small [7]. Two spin dissipation mechanisms have been suggested: (1) dissipation of spin angular momentum to the lattice, through spin-flip scattering from phonons or lattice defects [8]. Usually the contribution from lattice-defect-scattering is neglected and the demagnetization by optical laser pulses is described in terms of scattering from phonons [8]. (2) An alternative description is the purely electronic dissipation via non-local superdiffusive spin-currents [16-20]. In this case electronic relaxation processes are usually assumed to be spin-conserving altering the sample magnetization only locally. Although experimental evidence for both mechanisms has been reported, their relative contributions to ultrafast demagnetization remain debated with the accurate modelling of the fs laser-induced highly non-equilibrium state remaining a key obstacle.

Here we treat these controversial mechanisms of spin transport and scattering on equal footing. We utilize the recently demonstrated ability of single-cycle THz pulses to drive spin currents in metallic ferromagnets [21]. With the THz pulse duration of the order of the electronic and spin scattering events [21] it is possible to assess and accurately model the influence of elementary scattering processes on the sample magnetization while a non-equilibrium current is flowing in the magnetic material. This approach differs from the one usually taken in ultrafast optical demagnetization experiments where the accumulative nature of spin-flip scattering of an ensemble moving towards thermal equilibrium is measured [8].

In this letter we use single-cycle THz pulses with peak fields up to 60 MV/m to drive damage-free ultrafast spin currents in ferromagnetic thin films. We disentangle the influence of magnetic, $H_{THz}$, and electric, $E_{THz}$, field components of the THz pulse on the sample magnetization [22] in the time domain. The sample magnetization normal to the film surface is probed stroboscopically as a function of pump-probe delay using the polar magneto-optical Kerr effect (MOKE). During the THz pulse magnetization precession is observed due to a torque by $H_{THz}$. This precession is linear in $H_{THz}$ and the sense of the precession changes when the THz polarity is reversed. The precessing magnetization motion stops after the THz pulse has passed due to the single-cycle nature of the pulse and negligible energy dissipation on the sub-ps timescale [22–25]. In contrast we see a lingering sample demagnetization caused by THz-driven spin currents in the material. This effect is quadratic in the THz field strength and can be separated from the magnetization precession in a straightforward way. THz conductivity measurements finally allow us to relate these observations to defect-induced spin-lattice scattering processes of Elliot-Yafet type [8].

We use two samples with very different amount of defect sites: (a) a 9 nm thick epitaxial Fe thin film grown on a 500 μm thick MgO(001) substrate, capped with an ultrathin MgO layer, and (b) a 5 nm thick amorphous CoFeB film sputter deposited onto a silicon substrate. The full stack for the amorphous film, grown on 500 μm thick Si substrate, is $Al_2O_3(10)/CoFeB(5)/Al_2O_3$ (1.8), with thicknesses in nm. We characterized the THz conductivity of our samples by measuring the transmission of broadband THz radiation [21, 26] generated by a photoconductive switch pumped with 25 fs, 800 nm laser pulses from a 80 MHz laser oscillator. Amplitude and phase of the transmitted THz radiation are retrieved by means of electro-optical sampling in a ZnTe crystal. The THz radiation generated with this method is of relatively low intensity, with a peak field of less than 1 kV/m. We can reliably extract the optical constants for the magnetic films by normalizing to the transmission through identical but uncovered substrates [27].

Non-equilibrium spin dynamics was driven with the THz fields generated by optical rectification in a $LiNbO_3$ crystal using the 4 mJ, 100 fs, 800 nm central wavelength tilted wavefront pulses [28] from a 1 kHz regenerative amplifier. A small fraction of the laser pulses is picked off before the $LiNbO_3$ crystal, sent through a variable delay line, and used to probe the sample. Electro-optic sampling in a 100 μm thick GaP crystal shows that the electric field has the shape of a single-cycle transient [29] The Fourier spectrum of the pulses indicates that the bandwidth is approximately 2 THz with a 1 THz center frequency. Using Eq. (1) in ref. [30] we calculate the maximum peak electric field reached in this letter to 60 MV/m. Some measurements presented below were taken by reversing the direction of $H_{THz}$. This was achieved by inserting two polarizers into the THz beam limiting the peak electric field to 15 MV/m. We detect the magnetization state using the polar magneto-optical Kerr effect (MOKE) of the 800 nm probe pulse. The detection uses a balancing scheme with a Wollaston prism and two photodiodes.

The geometry of the THz pump - 800 nm MOKE probe experiments is depicted in Fig. 1. For both films, the static magnetization in the film plane was saturated along the y−direction with a 50 mT static magnetic field, larger than the coercivity field (1 mT for the Fe film, 5 mT for the CoFeB film). This configuration maximizes the torque between magnetization and THz magnetic field, given that the THz radiation is polarized so that its magnetic field component is along the x−direction. We also apply a larger external magnetic field $\mu_0 H_z$ = 0.6 T along the z−direction. This tilts the magnetization of the films out of the sample plane and allows for larger precession amplitudes, which are easier to detect in the polar MOKE geometry.

We first discuss the sample characterization in terms of their THz conductivity, σ. Figures 2(a) and (b) show real and imaginary parts of the THz conductivity for Fe and CoFeB samples, respectively. The crystalline Fe film in Fig. 2a shows a behavior that can be described well by the Drude model $\sigma(\omega) = \frac{\sigma_{DC}}{1-i\omega\tau}$, with $\sigma_{DC} = ne^2\tau/m$. Here $n$ is the carrier density, $e$ the electron charge and $m$ its mass. Fitting to the experimental data results in $\sigma_{DC} \approx 64$ kS/cm and a scattering time of $\tau = 30$ fs, close to the 100 kS/cm and 25 fs literature values for bulk Fe [31]. The behavior for the amorphous CoFeB sample shown in Fig. 2b is significantly different. First, the THz conductivity of CoFeB is about an order of magnitude smaller than that of the Fe film and it is suppressed at lower frequencies. Second, the imaginary part of the THz conductivity is negative. These experimental observations can be modeled using the Drude-Smith model [32]: $\sigma(\omega) = \frac{\sigma_{DC}}{1-i\omega\tau}\left(1 + \frac{C}{1-i\omega\tau}\right)$. It represents an extension of the standard Drude model where the parameter $C$, sometimes referred to as the persistence of velocity parameter, measures the backscattering probability at lattice defects and impurities ($-1 \leq C \leq 0$). $C = -1$ would describe a fully anisotropic backscattering of charge carriers, while $C = 0$ is the conventional Drude model with isotropic scattering. The parameter $C$ can hence be interpreted as the fraction of electrons that "bounce back" during a scattering event [33]. Fitting the data in Fig. 2b returns $\sigma_{DC} = 18$ kS/cm $\tau = 32$ fs, and a value $C \approx -0.7$, indicating substantial backscattering probability due to impurities or disorder in the system as expected for an amorphous CoFeB film. The conductivity data allow us also to estimate the skin depth $\delta = 2/\sigma \omega \mu_0$ for the two films. For both films, $\delta \sim 0.1 - 1$ μm, meaning that the current density, $J$, induced in the material by the THz electromagnetic field [21], is to a good approximation uniform across our films. These measurements, combined with transfer matrix calculations, also allow us to estimate the amount of energy deposited in the two films by the THz electromagnetic field. We find that approximately 15% of the incident intensity is absorbed in both films [29].

We now move on to discussing the magnetization dynamics induced by single-cycle THz pulses with high electromagnetic field strengths. Figures 3a and 3b show the measured sample magnetization response for both polarities of the THz field in Fe and CoFeB films, respectively. The static magnetization value is calculated comparing the data from vibrating sample magnetometry and static MOKE characterization (not shown). Figure 3a illustrates the response of the crystalline Fe film. At short time scales (up to ~ 2 ps) from the arrival of the THz pulse), the sample responds by preserving the phase of the THz pulse: upon sign reversal of the THz field, the magnetization dynamics also reverses its sign. After that, the system rapidly returns to the state before the arrival of the THz pulse.

Figure 3b shows the magnetization dynamics in the amorphous CoFeB film. At short time scales, this sample behaves very similar to Fe, with the magnetization's response changing sign upon reversal of the THz field polarity. However, the CoFeB sample does not return to its pristine state after the THz pulse has passed. At intermediate time scales (between 2ps and 10 ps), the magnetization settles to a level lower than the pre-pulse value, with no measurable dependence on the polarity of the THz signal. At even longer time scales, we observe oscillations of the magnetization consistent with the onset of the ferromagnetic resonance precession in the thin film [29].

In order to better understand and model the physics at play, we plot in Fig. 3c and 3d the difference and in Fig. 4a and 4b the sum of the data in Fig. 3a, and 3b taken for opposite THz polarities. The difference signal shown in Fig. 3c and 3d represents the sample magnetization component along the film normal, $M_z$, responding to the magnetic part of the THz field. It is accurately described by the Landau-Lifshitz equation $\frac{dM}{dt} = \gamma M x H$, where $\gamma$ = 28 GHz/T is the gyromagnetic ratio and H is the effective magnetic field, that comprises applied (including $H_{THz}$) and anisotropy fields. In equilibrium, M is aligned with the effective external magnetic field (excluding $H_{THz}$). While the THz pulse passes through the sample (for approximately the first 4 ps) $H_{THz}$ creates an additional torque that induces a precession of the magnetization [23]. For a small deviation of the magnetization from equilibrium, the Landau-Lifshitz equation has the analytical solution $M(t) = \gamma \sin\theta \int H_{THz}(t)dt$, where $\theta$ is the angle between M and H. In other words, the magnetization responds as the integral of the THz magnetic field, $H_{THz}(t)$ over time. This is demonstrated by the excellent agreement between the MOKE signal in Fig. 3a and 3b (symbols) with the numeric integral of the THz field (black solid line) measured by electro-optic sampling in GaP. The smaller extra peak in the THz field reference data at approximately 4 ps (dashed curve) arises from internal reflection within the 100 um thick GaP crystal, and it is therefore not present in the two magnetic samples grown on thicker substrates [29].

For the crystalline Fe sample, the Landau-Lifshitz equation is sufficient to fully describe the magnetization dynamics. In fact, as soon as the THz field leaves the sample, the magnetization relaxes back to its original direction, as no further time-resolved MOKE signal is detected down to the noise floor. The sum of the magnetization response for opposite THz polarity is shown in Fig. 4a and is zero within the sensitivity of our measurements over the whole time delay range. This can be understood from the fact that magnetic damping is simply not fast enough to facilitate energy dissipation out of the precessing spin system at such short times [22,23].

The situation is remarkably different for the amorphous CoFeB film, where a step-like response of the magnetization to the THz field is observed in the raw data of Fig. 3b. This is even more dominant in the the sum of the individual signals taken with opposite THz polarity shown in Fig. 4b. We identify this behavior as ultrafast demagnetization driven by the THz-induced current inside the material. This current is necessarily spin-polarized [21] since CoFeB is a ferromagnet. Figure 4c displays the THz peak field dependence of the demagnetization step function in Fig. 4b. The figure clearly shows that the demagnetization scales with the square of the THz peak field. Such a behavior is expected for energy dissipation due to scattering processes within a THz-driven spin current. In equilibrium this is responsible for Joule heating of the conductor that scales with $J \cdot E = \sigma E^2$, where $E$ is the internal electric field according to Ohm's law, $J = \sigma E$. In the following we model our measurements by the non-equilibrium analog of this dissipation process.

The spin current induced by the THz field inside a ferromagnetic film is uniform throughout the film as long as the film thickness is smaller than its skin depth as is the case here. The total energy dissipated by the THz electromagnetic field via electronic scattering processes is given by $\int J(t) \cdot E(t) dt \approx \sigma \int E(t)^2 dt$ as σ can be taken as being nearly constant in the $0.5 - 1.5$ THz frequency range, where most of the THz spectral density is found [29]. We corroborated this approximation by a full Fourier analysis including the finite dispersion of the conductivity plotted in Fig. 2b. We stress that $E(t)$ is the electric field inside the material. It is different in size to the incident THz electric field, $E_{THz}$, and its value and shape can be obtained from the magnetic response in Fig. 3d.

There are two possible dissipation channels for the THz-driven spin current. The dominant scattering channel is electronic scattering, conserving the total spin polarization of the material. It occurs with a characteristic scattering time of ~30 fs, as obtained by the THz conductivity measurements presented in Fig. 2. This is in good agreement with the average scattering times obtained for majority and minority spin carriers in ref. [21]. The second channel involves a change in the spin orientation of the scattered electrons. If the change in spin angular momentum remains within the electronic system it will not alter the total sample magnetization as detected by MOKE. However, spin-flip scattering can occur via the Elliot-Yafet mechanism that transfers the change in spin angular momentum to the lattice [8]. The energy dissipated by such spin-lattice scattering scales also as $\propto \int E(t)^2 dt$, as confirmed by the quadratic dependence of the demagnetization as a function of the THz field amplitude (Fig. 4c). This allows us to model the experimentally observed demagnetization ΔM in a compact form as $\Delta M \propto e^{-t/\tau_R} \int_{-\infty}^{t} E(\zeta)^2 d\zeta$, where the exponential term describes the recovery of the magnetization with time constant $\tau_R$ = 30 ps. The results are shown as the light blue line in Fig. 4b. It is important to note that the demagnetization data are matched by this model using only the size of the demagnetization as an adjustable parameter. In particular, we do not need to introduce any broadening of the fit to describe the demagnetization temporal response. This indicates that spin-lattice scattering timescales are very similar to that of spin conserving scattering events (~30 fs). Future experiments with faster THz field transients will allow us to determine this parameter even more precisely.

We now compare the observed THz-induced demagnetization with literature results. It is important to keep in mind the very different energy densities reached via fs optical laser and THz excitation. Following optical excitation the electronic system typically reaches electron temperatures above 1000 K corresponding to ~100 meV/atom [34]. In contrast we only reach typically ~ 0.01 meV/atom, as estimated by calculating the energy dissipation of a THz-driven spin current, $\int J(t) \cdot E(t) dt$, even for the highest THz field strengths used in this letter. It is, therefore, not surprising that for optical excitation the nature of the individual spin-lattice scattering events matters less than the relaxation of the highly excited non-equilibrium electronic system towards equilibrium. Optical demagnetization data are usually characterized by the demagnetization time, $\tau_M$, of the whole ensemble of spins [3,8]. For our Fe [6] and CoFeB [29] films we find $\tau_M$ ~ 100-200 fs in good agreement with expectations [8]. However, for our THz-driven demagnetization the individual spin-lattice scattering processes are far more relevant. We can, therefore, distinguish between spin-lattice scattering mediated by phonons and lattice defects. Demagnetization is only detected for defect-rich CoFeB and not for the near-perfect Fe single crystal films, even when the same amount of energy is deposited by the THz field. Our THz conductivity data in Fig. 2 point towards the strong influence of scattering from atomic disorder as a way to transfer spin angular momentum to the lattice. Future measurements will determine if a phonon-mediated spin-lattice transfer of angular momentum is non-existent or simply below the present detection limit in single-crystalline Fe films.

It is intriguing to take a further look at the energetics of defect-mediated spin-lattice scattering events. Electron-phonon coupling in general and spin-lattice scattering in particular require the excitation of lattice vibrations, possibly even localized at defect sites. We can estimate the average electron energy obtained by acceleration in the electric field, $E$, to an average speed, $v$, between scattering events as: $E\,v\,\tau$ ~ 0.01 meV. This indicates that only low-frequency phonons near the Brillouin zone center in Fe or CoFeB can be exited in individual scattering events. This may be the reason for the negligible spin-lattice scattering we observe in Fe, as electron-phonon coupling is typically faster for zone-boundary phonons [35]. The broken translational lattice symmetry near defect sites can lead to a far more efficient coupling to phonons explaining the increased spin-lattice scattering observed in Fig. 3 for amorphous CoFeB.

In conclusion, we demonstrated how THz-induced spin currents provide a novel tool to investigate the ultrafast transfer of spin angular momentum to the lattice. We find defect-mediated spin-lattice scattering processes to be surprisingly fast and to occur on similar timescales (~ 30 fs) than more conventional, spin-conserving scattering events. Our results are expected to stimulate new theoretical and experimental directions towards an encompassing and microscopic understanding of the physics of ultrafast demagnetization.

**Acknowledgments:** We are grateful to Tom Silva for useful discussion, to Aaron Lindenberg for support with the THz conductivity measurements and to Matthias Hudl for measuring the demagnetization induced by 800 nm pump light in CoFeB. This work is supported by the Department of Energy, Office of Science, Basic Energy Sciences, Materials Sciences and Engineering Division, under Contract DE-AC02-76SF00515. S.B. acknowledges support from the Swedish Research Council grant E0635001, and the Marie Sklodowska Curie Actions, Cofund, Project INCA 600398s.

References:

1. E. Beaurepaire, J.-C. Merle, A. Daunois, J.-Y. Bigot, Phys. Rev. Lett. **76**, 4250 (1996).

2. C. D. Stanciu, F. Hansteen, A. V. Kimel, A. Kirilyuk, A. Tsukamoto, A. Itoh, T. Rasing, Phys. Rev. Lett. **99**, 047601 (2007).

3. B. Koopmans, J. J. M. Ruigrok, F. Dalla Longa, W. J. M. De Jonge, Phys. Rev. Lett. **95**, 267207 (2005).

4. C. Stamm, *et al.* Nature Mat. **6**, 740 (2007).

5. F. Dalla Longa, J. T. Kohlhepp, W. J. M. De Jonge, B. Koopmans, Phys. Rev. B **75**, 224431 (2007).

6. E. Carpene, E. Mancini, C. Dallera, M. Brenna, E. Puppin, S. De Silvestri, Phys. Rev. B **78**, 174422 (2008).

7. B. Koopmans, M. Van Kampen, J. T. Kohlhepp, W. J. M. De Jonge, Phys. Rev. Lett. **85**, 844 (2000).

8. B. Koopmans, *et al.* Nature Mat. **9**, 259 (2010).

9. C. Boeglin, *et al.* Nature **465**, 458 (2010).

10. A. Kirilyuk, A. V. Kimel, and T. Rasing, Rev. Mod. Phys. **82**, 2731 (2010).

11. I. Radu, *et al.* Nature **472**, 205 (2011).

12. S. Mathias*, et al.* Proc. Natl. Acad. Sci. **109**, 4792 (2012).

13. S. Mangin, *et al.* Nature Mat. **13**, 286 (2014).

14. C.-H. Lambert, *et al.* Science **345**, 1337 (2014).

15. T. A. Ostler, *et al.* Nature Commun. **3**, 666 (2012).

16. G. Malinowski, *et al.* Nature Phys. **4**, 855 (2008).

17. M. Battiato, K. Carva, and P. M. Oppeneer, Phys. Rev. Lett. **105**, 027203 (2010).

18. K. Carva, M. Battiato, D. Legut, P. M. Oppeneer, Phys. Rev. B **87**, 184425 (2013).

19. D. Rudolf, *et al.* Nature Commun. **3**, 1037 (2012).

20. E. Turgut, *et al.* Phys. Rev. Lett. **110**, 197201 (2013).

21. Z. Jin, *et al.* Nature Phys. **11**, 761 (2015).

22. S. J. Gamble, *et al.* Phys. Rev. Lett. **102**, 217201 (2009).

23. C. Vicario, *et al.* Nature Photon. **7**, 720 (2013).

24. T. Kampfrath, *et al.* Nature Photon. **5**, 31 (2011).

25. A. H. M. Reid, T. Rasing, R. V. Pisarev, H. A. Dürr, M. C. Hoffmann, Appl. Phys. Lett. **106**, 082403 (2015).

26. P. U. Jepsen, D. G. Cooke, and M. Koch, Laser & Photonics Reviews **5**, 124 (2011).

27. L. Duvillaret, F. Garet, J.-L. Coutaz, Selected Topics in Quantum Electronics, IEEE Journal of 2, 739 (1996).

28. M. C. Hoffmann, J. A. Fülöp, J. Phys. D: Appl. Phys. **44**, 083001 (2011).

29. see online Supplemental Material.

30. M. C. Hoffmann, S. Schulz, S. Wesch, S. Wunderlich, A. Cavalleri, B. Schmidt, Opt. Lett. **36**, 4473 (2011).

31. M. A. Ordal, R. J. Bell, R. W. Alexander, L. A. Newquist, M. R. Querry, Appl. Opt. **27**, 1203 (1988).

32. N. V. Smith, Phys. Rev. B **64**, 155106 (2001).

33. R. Ulbricht, E. Hendry, J. Shan, T. F. Heinz, M. Bonn, Rev. Mod. Phys. **83**, 543 (2011).

34. H.-S. Rhie, H. A. Dürr, W. Eberhardt, Phys. Rev. Lett. **90**, 247201 (2003).

35. T. Chase, *et al.* Appl. Phys. Lett. **108**, 041909 (2016).

**Figures and captions**

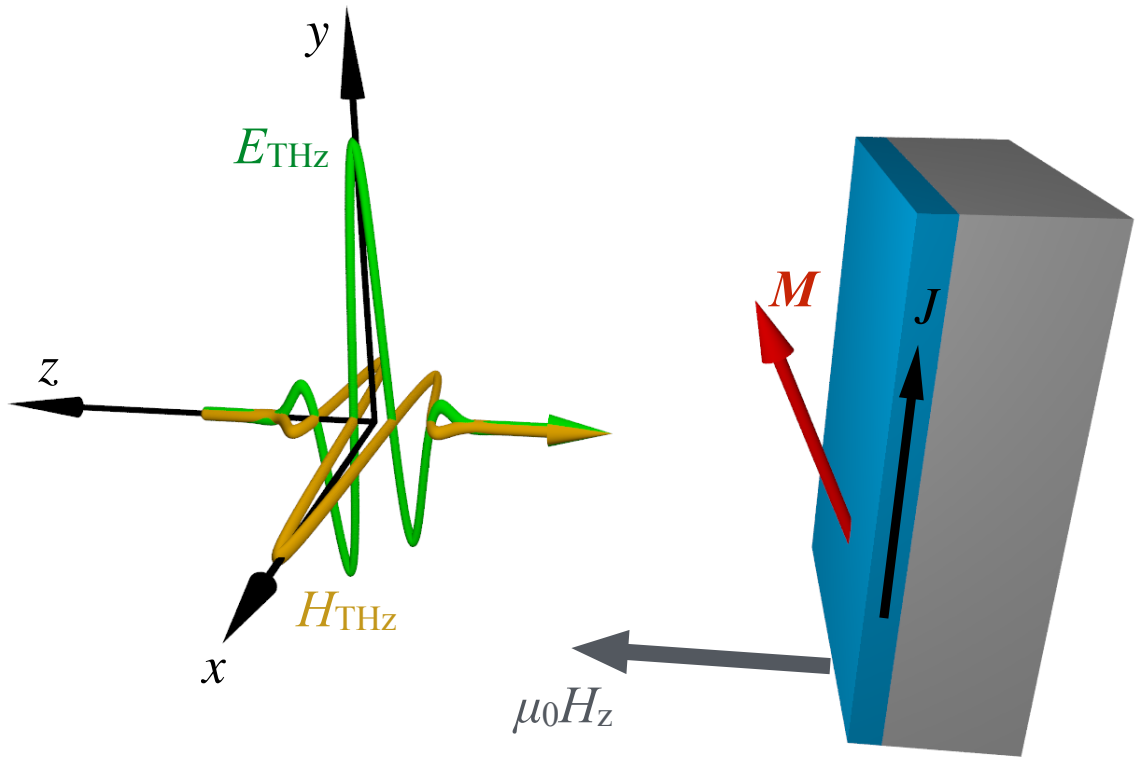

**Fig. 1**: Schematics of the experiment. The THz electric field, $E_{THz}$, is always polarized along the y axis, the THz magnetic field, $H_{THz}$, along the x axis. A static magnetic field is applied along the z-direction. An optical probe pulse (not shown) is incident collinearly with the THz pump pulse. It is used to measure the sample magnetization normal to the surface via the magneto-optical Kerr effect (MOKE) in reflection.

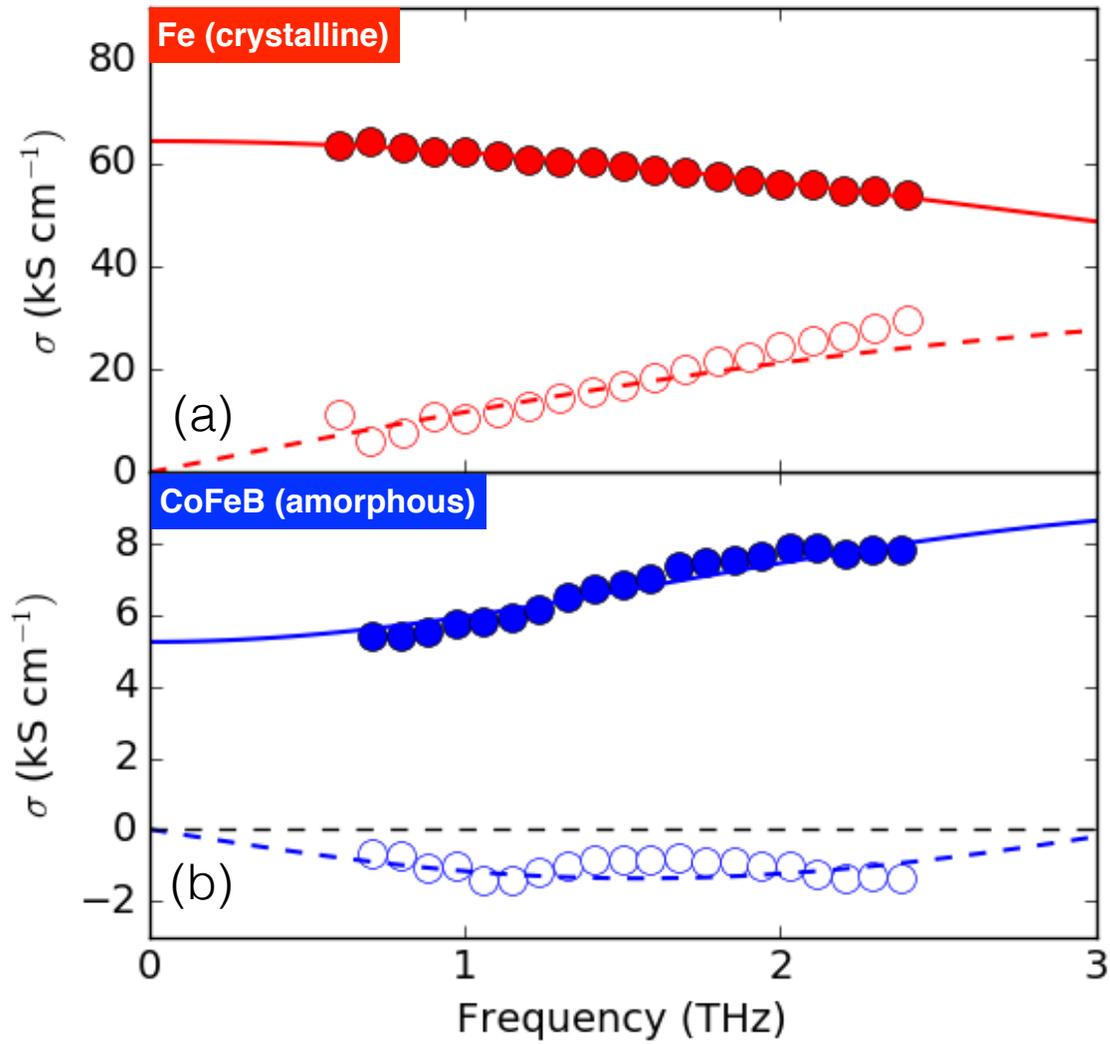

**Fig. 2:** Measured real (solid symbols) and imaginary (open symbols) parts of the frequency-dependent conductivity obtained from (a) crystalline Fe/MgO(001) and (b) amorphous CoFeB samples. The lines represent the Drude and Drude-Smith fitting to the experimental data in (a) and (b), respectively. The fitting parameters are given in the text.

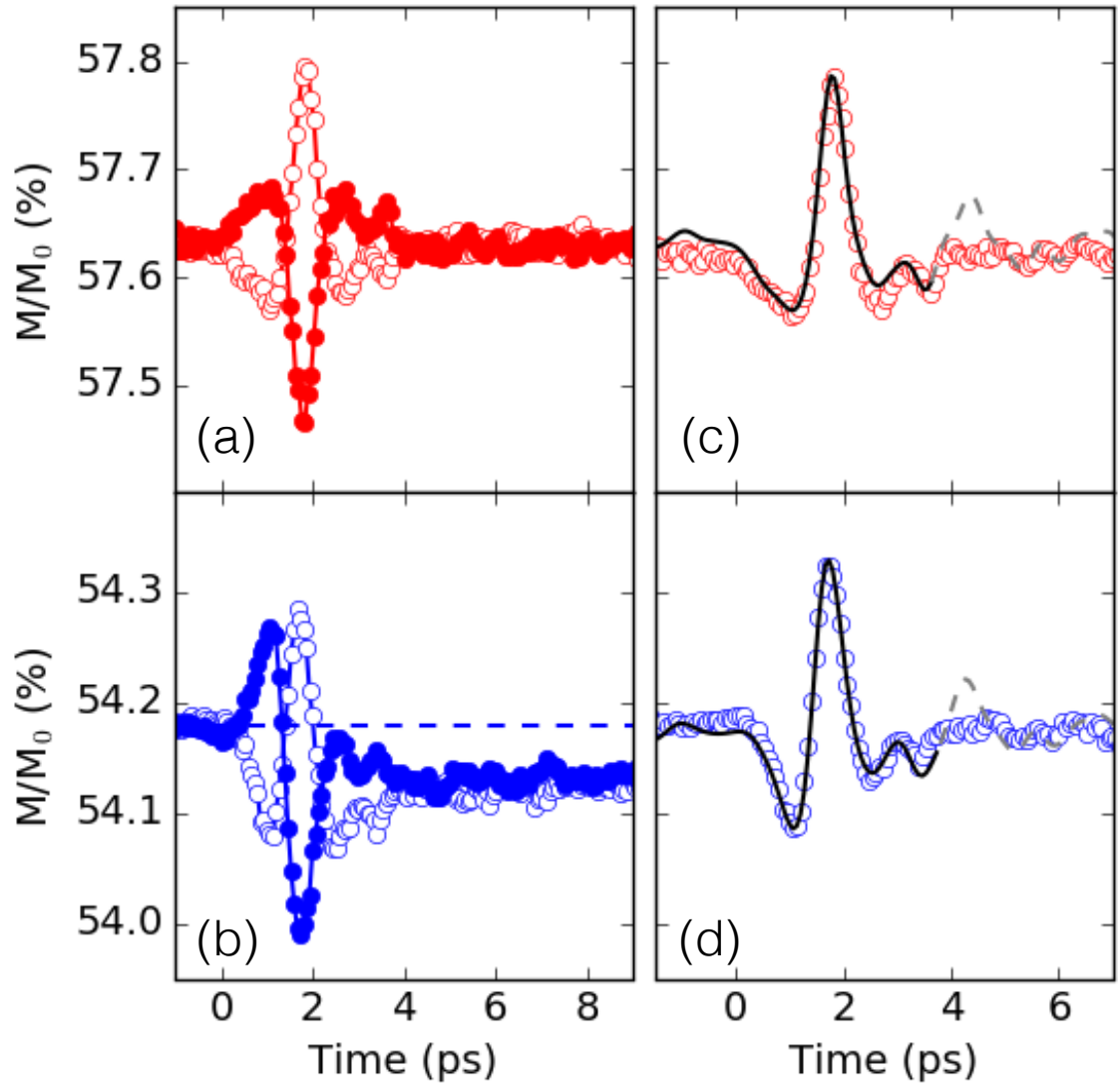

**Fig. 3:** Time-resolved magneto-optical Kerr effect (MOKE) response of the magnetization in (a) crystalline Fe and (b) amorphous CoFeB, for positive (open symbols) and negative (solid symbols) sign of the THz field. (c) and (d) Difference of the data in (a) and (b), respectively. The lines are the calculated magnetic response using the measured THz pulse shape as described in the text.

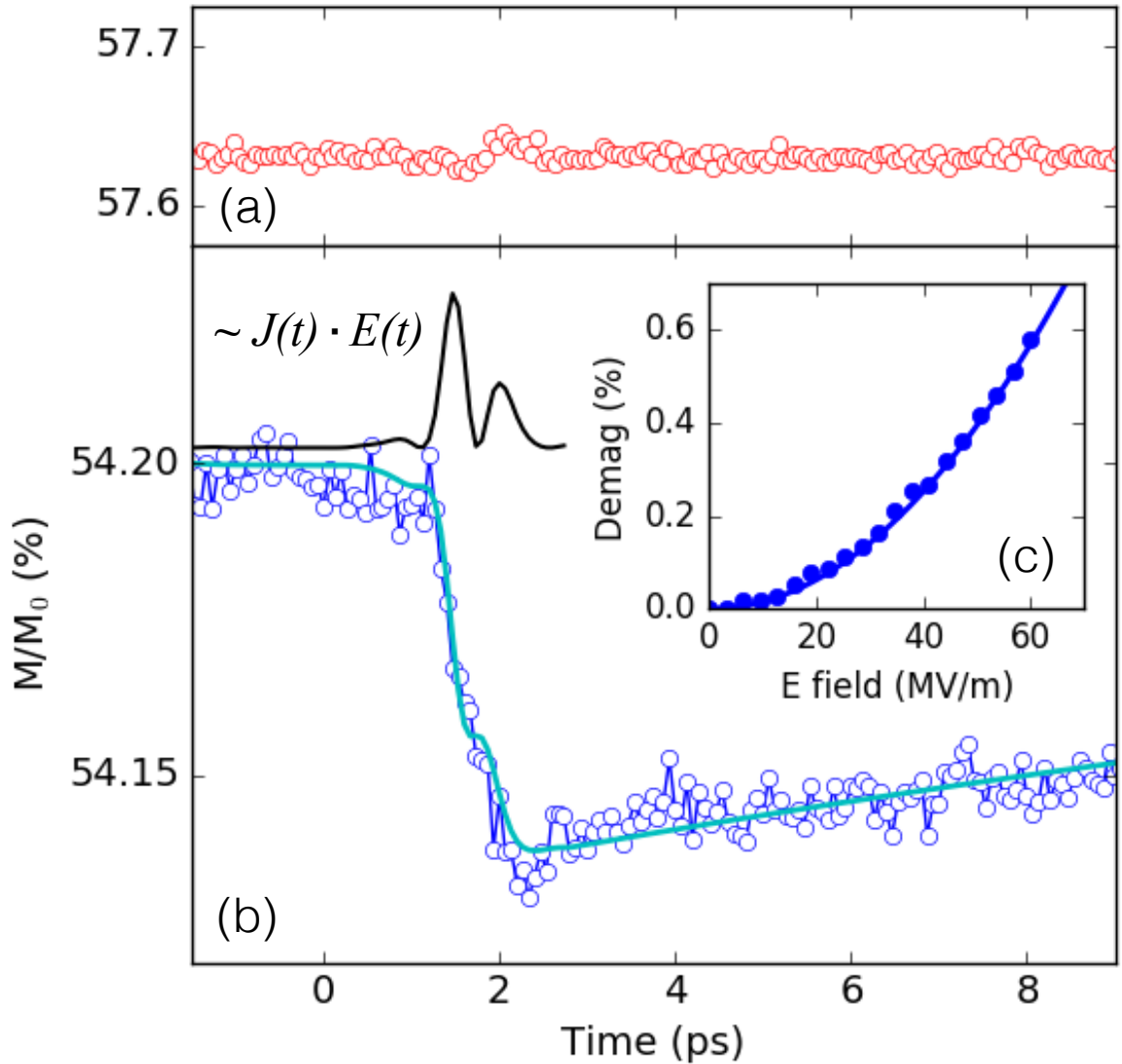

**Fig. 4**: Time-resolved magnetization dynamics following THz excitation. (a) and (b) represent the sum of the data shown in Fig. 3a and 3b for Fe and CoFeB, respectively. (c) THz peak field dependence of the CoFeB demagnetization data (solid symbols) in (b) and square fit (line). The cyan line in (b) is the integral over $J(t) \cdot E(t)$, with J being the THz driven current and E the THz electric field in the material, shown as black line in (b).